\documentclass[doublecol]{epl2} 
\usepackage{amssymb}
\newcommand{\half}{\frac{1}{2}}

\newcommand{\frth}{\frac{1}{4}}
\newcommand{\sxth}{\frac{1}{6}}

\newcommand{\z}{{\bar z}}
\title{Dark energy density predicted and explained}
\author{R. K. Nesbet}
\institute{
IBM Almaden Research Center, 
650 Harry Road,
San Jose, CA 95120-6099, USA}
\pacs{04.20.Cv}{Fundamental problems and general formalism}
\pacs{98.80.-k}{Cosmology}
\pacs{95.36.+x}{Dark energy}
\abstract{
It has recently been shown that the observed Hubble function for
cosmological expansion can be fitted accurately back to redshift unity
(7.33 Gyr ago) with only one free constant, while neglecting cosmic 
curvature and mass density, using the modified Friedmann equation 
implied by subjecting the Higgs scalar field model to conformal Weyl
scaling symmetry.  It is shown here that the implied dark energy
parameter is produced by dressing the bare scalar field by the 
neutral gauge boson field induced by weak cosmological time dependence 
of the conformal Higgs model.  Predicted persistent cosmic acceleration
is consistent with the nonclassical acceleration parameter inferred by 
conformal theory from observed excessive galactic rotation velocities 
and dark galactic halos, all without dark matter. 
}
\begin{document}
\maketitle
\section{Introduction}
\par In the currently accepted $\Lambda$CDM paradigm for cosmology,
dark energy $\Lambda$ remains without an explanation, while cold dark
matter CDM is assumed to be responsible for gravitational
phenomena that cannot be explained by general relativity as formulated
by Einstein.  The search for tangible dark matter has continued for
many years with no conclusive results\cite{SAN10}.
\par This situation motivates serious consideration of an alternative
paradigm. Universal conformal symmetry (local Weyl scaling covariance
\cite{WEY18}, for all massless elementary physical fields), promises 
a falsifiable alternative postulate\cite{MAN94,NES13}.
Conformal symmetry, already valid for fermion and gauge boson
fields\cite{DEW64}, is extended to both the metric tensor field of
general relativity and the Higgs scalar field of elementary-particle
theory\cite{HIG64,CAG98}.  This postulate is exemplified by conformal
gravity (CG)\cite{MAK89,MAN90,MAN06,MAN12} and by the conformal Higgs
model (CHM)\cite{NESM1} of cosmic Hubble expansion, without any novel 
elementary fields.
\par The fundamental postulate that all primitive fields have conformal
Weyl scaling symmetry is satisfied by spinor and gauge fields, but not
by the Higgs scalar field\cite{CAG98,MAN06}.
In the uniform, isotropic Robertson-Walker geometry of cosmic 
Hubble expansion, the Weyl tensor and Lagrangian density ${\cal L}_g$
of conformal gravity vanish identically\cite{MAN06}. To explain Hubble
expansion, the simplest assumption is existence of a conformal scalar 
field. The CHM retains the Higgs mechanism for gauge boson mass,
but acquires a gravitational effect described by a modified\cite{NESM1} 
CHM Friedmann cosmic evolution equation\cite{FRI22}, which replaces
Newton's gravitational constant by a Higgs field parameter of opposite 
sign\cite{MAN06,NESM1}.
\par A conformally invariant action integral is defined for complex 
scalar field $\Phi$ by the Lagrangian density\cite{MAN06,NESM1},
\begin{eqnarray}\label{LPhi0}
{\cal L}^0_\Phi=(\partial_\mu\Phi)^\dag\partial^\mu\Phi
-\sxth R\Phi^\dag\Phi.
\end{eqnarray}
Ricci scalar $R$ is trace $g_{\mu\nu}R^{\mu\nu}$ of the Ricci tensor.
\par  
Higgs parameters $w^2$ and $\lambda$ are defined by\cite{HIG64,CAG98} 
\begin{eqnarray}\label{DeltaL_par}
\Delta{\cal L}_\Phi=-V(\Phi^\dag\Phi)=
w^2\Phi^\dag\Phi-\lambda(\Phi^\dag\Phi)^2.
\end{eqnarray}
Omitting Ricci scalar $R$, parameters $w^2$ and $\lambda$ are positive
constants\cite{HIG64,CAG98}. Stationary action implies constant finite
$\Phi^\dag\Phi=\phi_0^2=w^2/2\lambda$.
$\hbar\Phi$ and $\hbar w$ are energies.
\par The CHM postulates\cite{NESM1} 
${\cal L}_\Phi={\cal L}^0_\Phi+\Delta{\cal L}_\Phi$
such that
\begin{eqnarray}\label{LPhi}
{\cal L}_\Phi=(\partial_\mu\Phi)^\dag\partial^\mu\Phi+
(w^2-\sxth R -\lambda\Phi^\dag\Phi)\Phi^\dag\Phi.
\end{eqnarray}
This differs from conformal ${\cal L}$ considered by Mannheim
\cite{MAN03} by including Higgs term $w^2\Phi^\dag\Phi$. Because 
$w^2\neq0$ breaks conformal symmetry, it must be produced dynamically,
verified below. Nonzero $\Phi^\dag\Phi$ is shown here to generate 
$w^2>0$. Biquadratic term $-\lambda(\Phi^\dag\Phi)^2$
retains conformal symmetry.  $\lambda<0$ is not excluded in conformal 
theory, and is implied by empirical cosmic expansion data.
The Higgs solution for $\Phi^\dag\Phi=\phi_0^2$\cite{HIG64} eliminates 
$\lambda$ from the CHM Friedmann equation, which acquires Higgs 
tachyonic mass parameter $w^2$ as dark energy\cite{NESM1}. 
\par Neglecting derivative terms quadratic in the Hubble constant,
$\phi_0^2=-\zeta/2\lambda$ implies dimensionless $\lambda<0$, 
where $\zeta(t)=\sxth R(t)-w^2(t)>0$\cite{NESM1}.
Given $\zeta$ and $\phi_0$ at present time $t_0$,
$\lambda(t_0)=-0.185\times 10 ^{-88}$.
\par The CHM Friedmann equation was integrated back to the earliest
cosmic time\cite{NESM1}.  Omitting dark matter and novel fields,
parameters were fitted to observed Hubble expansion data and to two
dimensionless quantities characteristic of the CMB, shift ratio 
$R(z)$\cite{KOM09} and acoustic scale $\ell_A(z)$\cite{WAM07} for 
redshift $z=z_*=1090$. Acceleration weight $\Omega_q=-q(t)$ was  
found to be positive since the earliest time.
\par The present paper is concerned with verifying theory underlying 
Ref\cite{NESM1}, in particular the origin of dark energy term $w^2$ in
the conformal Higgs model. Neglecting background mass and curvature, an
analytical solution of the CHM Friedmann equation, valid for $z\leq 1$,
replaces numerical integration\cite{NESM1} here. 
Table \ref{TabA1}, below, shows the high empirical accuracy 
of this solution back to redshift unity.  Nonzero mass/energy Friedmann 
weight $\Omega_m$ might be determined by recent expansion data for 
$z>1$\cite{RAL17}.
\par Augmenting the metric tensor field of general relativity by scalar
and vector fields has a long history, motivated variously by
unification with Maxwell's theory\cite{WEY18}, by Mach's principle
\cite{BAD61,DIC62}, by explaining Newton's constant\cite{ZEE79,SMO79}
in analogy to electroweak theory, and more recently to explain observed
anomalous galactic rotation\cite{MOF06,BEK04}.
The present theory differs from its predecessors by requiring both
gravitational and scalar bare field action integrals to be strictly
invariant under local Weyl scaling (conformal symmetry).  This
determines unique Lagrangian densities for both fields\cite{MAN06}.

\section{Review of variational theory}
\par Variational theory for fields in general relativity is a
straightforward generalization of classical field theory\cite{NES03}.
Given scalar Lagrangian density ${\cal L}=\sum_a{\cal L}_a$, action 
integral $I=\int d^4x \sqrt{-g} {\cal L}$ is required to be stationary 
for all differentiable field variations, subject to appropriate boundary
conditions.  $g$ here is the determinant of metric tensor $g_{\mu\nu}$.
Standard conservation laws follow from the variational principle. 
\par Gravitational field equations are determined by metric
functional derivative
$X^{\mu\nu}= \frac{1}{\sqrt{-g}}\frac{\delta I}{\delta g_{\mu\nu}}$.
Any scalar ${\cal L}_a$ determines energy-momentum tensor
$\Theta_a^{\mu\nu}=-2X_a^{\mu\nu}$, evaluated for a solution of the
coupled field equations.  Generalized Einstein equation
$\sum_aX_a^{\mu\nu}=0$ is expressed as
$X_g^{\mu\nu}=\half\sum_{a\neq g}\Theta_a^{\mu\nu}$.  Hence summed
trace $\sum_ag_{\mu\nu}\Theta_a^{\mu\nu}$ vanishes for exact field
solutions.  
Given $\delta{\cal L}=x^{\mu\nu}\delta g_{\mu\nu}$,
metric functional derivative
$\frac{1}{\sqrt{-g}}\frac{\delta I}{\delta g_{\mu\nu}}$ is
$X^{\mu\nu}=x^{\mu\nu}+\half{\cal L}g^{\mu\nu}$,
evaluated for a solution of the field equations.
Tensor $\Theta^{\mu\nu}=-2X^{\mu\nu}$ is symmetric.
\par For fixed coordinates $x^\mu$, local Weyl scaling is defined by
$g_{\mu\nu}(x)\to g_{\mu\nu}(x)\Omega^2(x)$\cite{WEY18} for arbitrary
real differentiable $\Omega(x)$. Conformal symmetry is defined by
invariant action integral $I=\int d^4x\sqrt{-g}{\cal L}$.
For any Riemannian tensor $T(x)$, $T(x)\to \Omega^d(x)T(x)+{\cal
R}(x)$ defines weight $d[T]$ and residue ${\cal R}[T]$. For a scalar
field, $\Phi(x)\to\Phi(x)\Omega^{-1}(x)$, so that $d[\Phi]=-1$.
Conformal Lagrangian density ${\cal L}$ must have weight
$d[{\cal L}]=-4$ and residue ${\cal R}[{\cal L}]=0$ up
to a 4-divergence\cite{MAN06}.

\section{Dark energy in the conformal Higgs model}
\par Because Higgs parameter $w^2$ breaks conformal symmetry, 
it must be of dynamical origin, as are gauge boson masses. 
The derivation here shows that conformal modification of the 
scalar field equation introduces time dependence on the scale of
the Hubble constant.   Finite Higgs field amplitude induces a weak
source current density for neutral gauge field  $Z_\mu$.  The induced
field dresses the scalar field, determining the $w^2$ term in its
effective Lagrangian density.  The conclusion is that dark energy 
is an unanticipated consequence of the Higgs mechanism.
\par A uniform, isotropic cosmos is described in cosmological theory
by Robertson-Walker (FLRW) metric
\begin{eqnarray}
ds^2=dt^2-a^2(t)(\frac{dr^2}{1-kr^2}+r^2d\omega^2),
\end{eqnarray}
where $c=\hbar=1$ and $d\omega^2=d\theta^2+\sin^2\theta d\phi^2$.
$k$ is a cosmic curvature constant.  Dimensionless scale factor 
$a(t)$, which determines Hubble expansion, satisfies a Friedmann 
equation\cite{FRI22}.  Redshift $z(t)=1/a(t)-1$.
\par Log derivatives of the Higgs field, on the order of Hubble
constant $H_0$, are very small.  Neglecting terms quadratic in $H_0$,
the effective scalar field equation is
$\frac{\delta{\cal L}_\Phi}{\delta\Phi^\dag}=
(w^2-\sxth R-2\lambda\Phi^\dag\Phi)\Phi=0$, implying
$\Phi^\dag\Phi=\phi_0^2=(w^2-\sxth R)/2\lambda$. 
For this value of $\phi_0^2$,
${\cal L}_\Phi=\half(w^2-\sxth R)\phi_0^2$
\par Metric functional derivative
$\frac{1}{\sqrt{-g}}\frac{\delta I_\Phi}{\delta g_{\mu\nu}}$ is
$X_\Phi^{\mu\nu}=x_\Phi^{\mu\nu}+\half{\cal L}_\Phi g^{\mu\nu}$, where
$x_\Phi^{\mu\nu}=\sxth R^{\mu\nu}\phi_0^2$. 
This implies $X_\Phi^{\mu\nu}=
\sxth\phi_0^2(R^{\mu\nu}-\frth Rg^{\mu\nu}+\frac{3}{2}w^2g^{\mu\nu})$,
which produces a modified gravitational equation\cite{NESM1},
\begin{eqnarray} \label{greqn}
 R^{\mu\nu}-\frth Rg^{\mu\nu}+{\bar\Lambda}g^{\mu\nu}
 =-{\bar\tau}\Theta_m^{\mu\nu}.
\end{eqnarray}
Parameters here are ${\bar\Lambda}=\frac{3}{2}w^2$ and
${\bar\tau}=-3y^2/\phi_0^2$, replacing Einstein $\tau=8\pi G/c^4$.  
Free radiation energy-momentum is included in $\Theta_m^{\mu\nu}$.  
\par Numerical factor $y^2$, to be determined from empirical data, 
allows for a dimensionless coefficient of conformal ${\cal L}_\Phi$.
The reversed sign of the effective gravitational constant in the
conformal Higgs model gives a radically different qualitative picture
of early cosmic expansion\cite{NESM1,NES13}.  Primordial mass/energy
density would by itself cause accelerated cosmic expansion.
\par Ricci tensor $R^{\mu\nu}$ for the FLRW metric depends on
$a(t)$ through two independent functions,
$\xi_0(t)=\frac{\ddot a}{a}$ and
$\xi_1(t)=\frac{{\dot a}^2}{a^2}+\frac{k}{a^2}$,
such that $R^{00}=3\xi_0$ and scalar $R=6(\xi_0+\xi_1)$.
The CHM Friedmann equation 
\begin{eqnarray} \label{CFeq}
-\frac{2}{3}(R^{00}-\frth Rg^{00})= \xi_1(t)-\xi_0(t)=
\nonumber\\
\frac{{\dot a}^2}{a^2}+\frac{k}{a^2}-\frac{\ddot a}{a}
    =\frac{2}{3}({\bar\tau}\rho+{\bar\Lambda}), 
\end{eqnarray}
given energy density $\rho=\Theta_m^{00}$, determines FLRW scale 
parameter $a(t)$ and Hubble function
$H(t)=\frac{{\dot a}}{a}$\cite{NESM1}. Hubble constant $H_0=H(t_0)$ at
present time $t_0$.  Vanishing trace in conformal theory  eliminates 
one of the two independent Friedmann equations of standard theory.  
\par Defining $\frac{{\dot a}}{a}=h(t)H_0$, it is convenient to use 
Hubble units, such that $c=\hbar=1$, and dimensionless
$a(t_0)=1$, $h(t_0)=1$. The units for frequency, energy, and 
acceleration are $H_0,\hbar H_0,cH_0$, respectively.
\par Dividing the CHM Friedmann equation by 
$\frac{{\dot a}^2}{a^2}(t)=H^2(t)$  
gives dimensionless sum rule\\
$\Omega_m(t)+\Omega_\Lambda(t)+\Omega_k(t)+\Omega_q(t)=1$, where\\
$\Omega_m(t)= \frac{2}{3}\frac{{\bar\tau}\rho(t)}{H^2(t)}<0$,
$\Omega_\Lambda(t)=\frac{2}{3}\frac{{\bar\Lambda}}{H^2(t)}
 =\frac{w^2}{H^2(t)}>0$,
$\Omega_k(t)=\frac{-k}{a^2(t)H^2(t)}$, and
$\Omega_q(t)=\frac{{\ddot a}a}{{\dot a}^2}=-q(t)$.
$\rho$ here includes both mass and radiation energy density.
\par Cosmological constant ${\bar\Lambda}=\frac{3}{2}w^2>0$ is
consistent with electroweak theory.  For positive energy density
$\rho$, ${\bar\tau}\rho<0$, as shown by Mannheim\cite{MAN03}.
\par Mannheim\cite{MAN03} has fitted observed type Ia supernovae
luminosities for redshifts $z\leq1$ to a formula based
on the standard Friedmann equation, with $\Omega_m=0$, within the
statistical error of empirical luminosity distances $d_L$:
\begin{eqnarray} \label{MANdL}
\frac{H_0d_L}{c}=\frac{(1+z)^2}{-q_0}(1-[1+q_0
-\frac{q_0}{(1+z)^2}]^\half),
\end{eqnarray}
evaluated for $-q_0=\Omega_q(t_0)=\Omega_\Lambda(t_0)=0.37$.
\par For $\Omega_m=0$, the standard Friedmann sum rule requires
$\Omega_k=1-\Omega_\Lambda$, implying $\Omega_k$ much larger than
empirically anticipated\cite{KOM09}.
It is shown below that CHM Friedmann Eq.(\ref{CFeq}) fits
the same data with both $\Omega_m$ and $\Omega_k$ set to zero.
The CHM sum rule reduces to 
$\Omega_\Lambda+\Omega_q=1$\cite{NESM1},
which determines acceleration weight $\Omega_q>0$.

\section{Fit to observed Hubble expansion}
\par In Hubble units $t(HUB)=t(MKS)H_0(MKS)$ and
$\frac{{\dot a}}{a}(t)=h(t)$ are dimensionless.
Setting $\Omega_k=\Omega_m=0$, for 
$\alpha=w^2=\Omega_\Lambda(t_0)$, the CHM Friedmann 
equation\cite{NESM1,NES13} is
\begin{eqnarray} \label{cFeq}
 \frac{{\dot a}^2}{a^2}-\frac{\ddot a}{a}=
    \frac{2}{3}{\bar\Lambda}=\alpha.
\end{eqnarray}
\par In Hubble units, Eq.(\ref{cFeq}) reduces to 
$\frac{d}{dt}h(t)=-\alpha$.
The explicit solution for $t\leq t_0$ is
\begin{eqnarray}
h(t)=\frac{{\dot a}}{a}(t)=\frac{d}{dt}\ln a(t)=1+\alpha(t_0-t),
\nonumber\\
\ln a(t)=-(t_0-t)-\half\alpha(t_0-t)^2,
\nonumber\\
a(t)=\exp[-(t_0-t)-\half\alpha(t_0-t)^2].
\end{eqnarray}
\par From the definition of redshift $z(t)$: \\
$1+z(t)=\frac{1}{a(t)}=\exp[(t_0-t)+\half\alpha(t_0-t)^2],\\
(t_0-t)^2+\frac{2}{\alpha}(t_0-t)=\frac{2}{\alpha}\ln(1+z)$.
\par In Hubble units with $\frac{{\dot a}}{a}(t_0)=1$, this implies
\begin{eqnarray}   \label{dteq}
t_0-t(z)=(\sqrt{2\alpha\ln(1+z)+1}-1)/\alpha,
\nonumber\\
\frac{dt}{dz}=\frac{-1}{(1+z)\sqrt{2\alpha\ln(1+z)+1}}.
\end{eqnarray}
If $\alpha=0.732$ and $z=1$, $(t_0-t)/H_0=7.33Gyr$, for
$H_0=67.8\pm 0.9 km/s/Mpc=2.197\times 10^{-18}/s$\cite{PLC15}.
\par Neglecting both curvature weight $\Omega_k$ and cosmic mass/energy
weight $\Omega_m$, conformal sum rule
$\Omega_\Lambda+\Omega_q=1$ holds for acceleration weight
$\Omega_q=\frac{{\ddot a}a}{{\dot a}^2}$.
For $\Omega_k=0$, luminosity distance $d_L(z)=(1+z)\chi(z)$, where
\begin{eqnarray} \label{Chi_z}
\chi(z)=
\int_{t(z)}^{t_0}\frac{dt}{a(t)}=
 \int_z^0 d\z(1+\z)\frac{dt}{dz}(\z)=
\nonumber\\
 \int_0^z\frac{d\z}{\sqrt{2\alpha\ln(1+\z)+1}}.
\end{eqnarray}
\par Evaluated for parameter $\alpha=\Omega_\Lambda(t_0)=0.732$,
the fit to scaled luminosity distances $H_0d_L/c$ from Hubble
expansion data\cite{MAN03,NESM1} is given in Table \ref{TabA1}.
\begin{table}[h]
\caption{Scaled luminosity distance fit to Hubble data} \label{TabA1}
\begin{tabular}{lcccc}
 &                 &          &Theory     &Observed    \\
z& $\Omega_\Lambda$&$\Omega_q$&$H_0d_L/c$
 &$H_0d_L/c$\cite{MAN03}\\ \hline
0.0& 0.732& 0.268& 0.0000& 0.0000\\
0.2& 0.578& 0.422& 0.2254& 0.2265\\
0.4& 0.490& 0.510& 0.5013& 0.5039\\
0.6& 0.434& 0.566& 0.8267& 0.8297\\
0.8& 0.393& 0.607& 1.2003& 1.2026\\
1.0& 0.363& 0.637& 1.6209& 1.6216
\end{tabular}
\end{table}
\par Because conformal gravitational ${\cal L}_g$ vanishes identically
in uniform isotropic geometry\cite{MAN06}, Hubble expansion is driven
by the gravitational field equation due to $-\sxth R\Phi^\dag\Phi$
in conformal ${\cal L}_\Phi$\cite{NESM1}.  Higgs parameters
$w^2,\lambda$ determine Ricci scalar $R$ and a cosmological
constant\cite{NESM1}.

\section{Conformal scalar field}
\par The conformal scalar field equation
including parametrized $\Delta{\cal L}_\Phi$ is\cite{MAN06,NESM1}
\begin{eqnarray}\label{Phieq}
\frac{1}{\sqrt{-g}}\partial_\mu(\sqrt{-g}\partial^\mu\Phi)=
 (-\sxth R+w^2-2\lambda\Phi^\dag\Phi)\Phi.
\end{eqnarray}
Ricci scalar $R$ introduces gravitational effects.
\par Only real-valued solution $\phi(t)$ is relevant in uniform,
isotropic geometry. Defining $V(\phi)=
 \half(\zeta+\lambda\phi^2)\phi^2$,
where $\zeta(t)=\sxth R-w^2$, the field equation is
\begin{eqnarray} \label{phieq}
\frac{{\ddot\phi}}{\phi}
+3\frac{{\dot a}}{a}\frac{{\dot\phi}}{\phi}=
 -(\zeta(t)+2\lambda\phi^2).
\end{eqnarray}
\par Omitting $R$ and assuming constant $\lambda>0$ and $w^2$, Higgs
solution $\phi_0^2=w^2/2\lambda$\cite{HIG64} is exact.
All time derivatives drop out.  In the conformal scalar field equation,
cosmological time dependence of Ricci scalar $R(t)$, 
determined by the CHM Friedmann cosmic evolution equation,
introduces nonvanishing time derivatives and implies $\lambda<0$.

\section{Determination of parameter $w^2$}
\par The Higgs model\cite{HIG64,CAG98} derives gauge boson mass from 
coupling via gauge covariant derivatives to a postulated SU(2) doublet 
scalar field $\Phi$.  SU(2) symmetry is broken by a solution of the 
field equation such that $\Phi^\dag\Phi=\phi^2(t)$ where $\phi=\phi_0$, 
constant in space.
Only the neutral component of charge-doublet field $\Phi$ is nonzero.
\par An essential result of the Higgs model, generation of gauge
field masses, follows from a simplified semiclassical theory of the
coupled scalar and gauge fields\cite{HIG64}.  This is extended here to
include the gravitational field, metric tensor $g_{\mu\nu}$, but 
greatly simplified by assuming the standard cosmological model
described by Robertson-Walker geometry.  This is further simplified,
for the purpose of establishing credibility and orders of magnitude,
by considering only the neutral vector field $Z^\mu$.  Numerical 
results follow from solving nonlinear coupled field equations.
\par Gauge invariance replaces bare derivative $\partial_\mu$
by gauge covariant derivative\cite{CAG98}
\begin{eqnarray}
D_\mu=\partial_\mu-\frac{i}{2}g_z Z_\mu.
\end{eqnarray}
This retains ${\cal L}_Z$ in terms of
$Z_{\mu\nu}=\partial_\mu Z_\nu-\partial_\nu Z_\mu$ 
and augments conformal 
\begin{eqnarray}
{\cal L}^0_\Phi=
(\partial_\mu\Phi)^\dag \partial^\mu\Phi-\sxth R\Phi^\dag\Phi 
\end{eqnarray} 
by coupling term 
\begin{eqnarray} \label{DelD}
\Delta{\cal L}=
(D_\mu\Phi)^\dag D^\mu\Phi -(\partial_\mu\Phi)^\dag \partial^\mu\Phi=
\nonumber\\
 \frac{i}{2}g_z Z_\mu^*\Phi^\dag\partial^\mu\Phi
-\frac{i}{2}g_z Z^\mu(\partial_\mu\Phi)^\dag\Phi
+\frth g_z^2\Phi^\dag Z_\mu^* Z^\mu\Phi.
\end{eqnarray}
\par Parametrized for a generic complex vector field\cite{CAG98}, 
\begin{eqnarray} \label{DelZ}
\Delta{\cal L}_Z=
 \half m^2_Z Z^*_\mu Z^\mu-\half(Z^*_\mu J_Z^\mu+Z^\mu J^*_{Z\mu}),
\end{eqnarray}
given mass parameter $m_Z$ and source current density $J_Z^\mu$.
The field equation for parametrized $Z^\mu$ is\cite{CAG98}
\begin{eqnarray}\label{Zeq}
 \partial_\nu Z^{\mu\nu}=
\frac{2}{\sqrt{-g}}\frac{\delta\Delta I_Z}{\delta Z_\mu^*}=
m_Z^2 Z^\mu-J_Z^\mu . 
\end{eqnarray} 
$\Delta{\cal L}$ from $D_\mu$, Eq.(\ref{DelD}), 
determines parameters for field $Z^\mu$:
\begin{eqnarray}\label{Zcdeq}
 \frac{2}{\sqrt{-g}}\frac{\delta\Delta I}{\delta Z_\mu^*}=
 \half g_z^2\Phi^\dag\Phi Z^\mu+ig_z\Phi^\dag\partial^\mu\Phi 
\end{eqnarray}
implies not only Higgs mass formula $m_Z^2=\half g_z^2\Phi^\dag\Phi$,
but also field source density
$ J_Z^\mu=-ig_z\Phi^\dag\partial^\mu\Phi$.
Time variation of real field $\phi$ determines pure imaginary 
$J^0_Z=-ig_z\phi\partial^0\phi
 =-ig_z\frac{{\dot\phi}}{\phi}\phi^2$.
Cosmological time variation of Ricci scalar $R$ implies nonvanishing
real parameter $\frac{{\dot\phi}}{\phi}$.
\par The same logic can be applied to scalar field $\Phi$. Functions 
of gauge fields in the gauge covariant derivative of $\Phi$
can be identified with otherwise arbitrary parameters in the model
$\Delta{\cal L}_\Phi$ of Eq.(\ref{DeltaL_par}).  Defining  
$\Delta I_\Phi=\int d^4x\sqrt{-g}\Delta{\cal L}_\Phi$, the parametrized
effective potential term in the  scalar field equation is given by
Eq.(\ref{DeltaL_par}):
\begin{eqnarray} \label{DPhieq}
\frac{1}{\sqrt{-g}}\frac{\delta\Delta I_\Phi}{\delta\Phi^\dag}=
 (w^2-2\lambda\Phi^\dag\Phi)\Phi . 
\end{eqnarray}
Using $\Delta{\cal L}$ derived from the covariant derivative,
\begin{eqnarray}
 \frac{1}{\sqrt{-g}}\frac{\delta\Delta I}{\delta \Phi^\dag}=
\frth g_z^2Z_\mu^*Z^\mu \Phi
+\frac{i}{2}g_z(Z_\mu^*+Z_\mu)\partial^\mu\Phi.
\end{eqnarray}
Comparison with Eq.(\ref{DPhieq}) implies 
$w^2=\frth g_z^2 Z_\mu^*Z^\mu$.
Neglecting derivatives of induced gauge field $Z^\mu$,
Eq.(\ref{Zeq}) reduces to $Z^\mu= J_Z^\mu/m_Z^2 $.
Implied pure imaginary $Z^\mu$ does not affect parameter $\lambda$.
However, $|Z^0|^2=\frac{4}{g_z^2}(\frac{{\dot\phi}}{\phi})^2$,
so that the scalar field equation implies nonvanishing 
$w^2=\frth g_z^2|Z^0|^2=(\frac{{\dot\phi}}{\phi})^2$.
\par Dividing ${\cal L}_\Phi$, $\Delta{\cal L}_\Phi$, and the implied
field equation by arbitrary constant $y^2$ does not alter the formula 
for $\phi^2$.
For $\Delta{\cal L}$ derived from 
covariant derivative $D_\mu$, Eq.(\ref{DelD}), 
\begin{eqnarray}
 \frac{2}{\sqrt{-g}}\frac{\delta\Delta I}{\delta Z_\mu^*}=
\frac{ g_z^2}{2y^2}\Phi^\dag\Phi Z^\mu
+\frac{ig_z}{y^2}\Phi^\dag\partial^\mu\Phi.
\end{eqnarray}
Eqs.(\ref{Zeq}) and (\ref{Zcdeq}) imply
$m_Z^2=\frac{g_z^2}{2y^2}\Phi^\dag\Phi$ and 
$J_Z^0=\frac{-ig_z}{y^2}\frac{{\dot\phi}}{\phi}\Phi^\dag\Phi$.
$Z^0=\frac{J_Z^0}{m_Z^2}=\frac{-2i}{g_z}\frac{{\dot\phi}}{\phi}$
cancels factor $y^2$ from 
$w^2=\frth g_z^2 Z^*_\mu Z^\mu=(\frac{{\dot\phi}}{\phi})^2$.
For postulated 
${\cal L}_\Phi/y^2$, modified Einstein equation
$X_\Phi^{\mu\nu}=\frac{y^2}{2}\Theta_m^{\mu\nu}$
implies parameter ${\bar\tau}=-3y^2/\phi^2$ in the conformal 
Friedmann equation.
\par Weyl scaling residues  
cancel exactly for real gauge fields\cite{MAN11}, so that the total 
energy-momentum tensor is  conformal and traceless. 
The trace condition can be verified
for the pure imaginary gauge field derived here\cite{NES13}.
\par Derivatives due to cosmological time dependence act as an extremely
weak perturbation of the Higgs scalar field.  The scalar field is 
dressed by an induced gauge field amplitude. Derivatives of the induced
gauge field (but not of $\Phi$) can be neglected and are omitted here.
\par The standard Higgs model omits $R$ and assumes constants
$w^2,\lambda>0$.
For trial solution $\Phi=\phi_0$ of the scalar field equation,
constant $\Phi^\dag\Phi=\phi_0^2=w^2/2\lambda$.  This implies
$\frac{{\dot\phi}_0}{\phi_0}=0$, which does not couple the fields.
\par In conformal theory Ricci scalar $R(t)$ varies in cosmological 
time, but $\zeta=\sxth R-w^2 >0$\cite{NESM1}.  Time-dependent 
$\phi_0^2(t)=\zeta(t)/(-2\lambda)$ is defined if $\lambda<0$,
but is no longer an exact solution unless Ricci $R$ is constant.
This results in small but nonvanishing $\frac{{\dot\phi}}{\phi}$, 
hence nonzero source density $J_Z^0$.  As discussed below, the coupled
field theory does not rule out negative $\lambda$.
\par Theory omitting charged gauge fields $W^\pm_\mu$ does not 
determine parameter $\lambda$.  $w^2$ and $\phi_0$ are well defined
if $\lambda<0$, as required by empirical $\sxth R>w^2$\cite{NESM1}.
Confirming the standard Higgs model, gauge field $Z_\mu$ acquires mass
$m_Z$ from coupling to $\Phi$.
\par The coupled fields break conformal and SU(2) symmetries.
Time-dependent $R$ implies nonvanishing real ${\dot\phi}$ and pure 
imaginary $J_Z^0$.  Complex solutions of the gauge field equations, 
induced by pure imaginary current densities, exist but do not 
preserve gauge symmetry. Imaginary gauge field amplitudes model quantum 
creation and annihilation operators.  Only the squared magnitudes of 
these nonclassical entities have classical analogs.

\section{Current values of time derivatives}
\par The accurate empirical fit of Table(\ref{TabA1}) assumes
constant $\alpha=w^2=\frac{{\dot\phi}^2}{\phi^2}$.  
Derivative terms of second order in $H_0$ are neglected in
approximate Higgs solution $\phi_0^2(t)=-\zeta(t)/2\lambda$,
where $\zeta(t)=\sxth R(t)-w^2$.
Weight parameters in Table(\ref{TabA1})
$\Omega_\Lambda=w^2a^2/{\dot a}^2, \Omega_q= {\ddot a}a/{\dot a}^2$
at present time $t_0$ are $\Omega_\Lambda=0.732, \Omega_q= 0.268$,
fitted to Hubble expansion data for redshifts $z\leq1$\cite{NESM1},
with $\Omega_m=0$. 
Time-dependent Ricci scalar $R=6(\xi_0(t)+\xi_1(t))$, where
$\xi_0(t)=\frac{\ddot a}{a}$ and
$\xi_1(t)=\frac{{\dot a}^2}{a^2}+\frac{k}{a^2}$\cite{NESM1}. 
For $k=0$, $\sxth R(t)=
 h^2(t)(2-\Omega_\Lambda(t)-\Omega_m(t))=
 h^2(t)(1+\Omega_q(t))>h^2(t)\Omega_\Lambda=w^2$.
\par $\zeta>0$ for computed $R(t)$\cite{NESM1} implies $\lambda<0$.
$\hbar\phi(t_0)=174GeV$\cite{AMS08}$=1.203\times 10^{44}\hbar H_0$ 
in Hubble units. For $\Omega_m=0$, $\zeta(t_0)=2\Omega_q(t_0)=0.536$ . 
If $\lambda(t)=\zeta/(-2\phi^2)$, $\phi(t)=\phi_0(t)$. Given
$\zeta(t_0)$ and $\phi(t_0)$, $\lambda(t_0)=-0.185\times 10^{-88}$. 
\par $\frac{{\dot\phi}}{\phi}$ can be estimated from computed
$\zeta(t)$.
Defining $\phi(t)=b(t)\phi_0(t)$, Eq.(\ref{phieq}) becomes
\begin{eqnarray} \label{bphieq}
\frac{{\ddot\phi}}{\phi}+3\frac{{\dot a}}{a}\frac{{\dot\phi}}{\phi}=
-(\zeta(t)+2\lambda\phi^2)=\zeta(b^2-1).
\end{eqnarray}
This can be integrated for $b(t)$ given $\lambda(t)$ and 
$\phi_0^2(t)=-\zeta(t)/2\lambda$ determined by the Friedmann equation.
\par Assuming $\phi\simeq\phi_0$ simplifies Eq.(\ref{bphieq}).
Given $\phi^2(t)=\zeta(t)/2\tau(t)$, where $\tau=|\lambda|$,
$2\frac{{\dot\phi}}{\phi}=
 \frac{{\dot\zeta}}{\zeta}-\frac{{\dot\tau}}{\tau}$.  Neglecting
$\Omega_m$ for $z\leq 1$, for $\alpha=\Omega_\Lambda(t_0)$,
$\zeta(t)=2h^2(t)\Omega_q(t)=2(h^2(t)-\alpha)$ 
implies $\zeta(t_0)=2(1-\alpha)=0.536$.
From Friedmann equation ${\dot h}=-\alpha$, 
${\dot\zeta}(t_0)=4h{\dot h}=-4\alpha=-2.928$. 
Then $\frac{{\dot\zeta}}{\zeta}(t_0)=-5.462$ in Hubble units.
$\frac{{\dot\phi}}{\phi}(t_0)=-\sqrt{\alpha}=-0.856$
implies $\frac{{\dot\tau}}{\tau}(t_0)=-3.750$.
Constant $\lambda$ implies $\frac{{\dot\phi}_0}{\phi_0}(t_0)=-2.731$.
Time-dependent $\lambda$ is consistent with a dynamical mechanism.

\section{Higgs model parameter $\lambda$} 
\par It is widely assumed that negative $\lambda$ would imply an
unstable physical vacuum, but the present analysis does not support
this conclusion.  A simple harmonic oscillator model is not valid
for the coupled field equations considered here.
At a more fundamental level, the field equations result from a
variational principle of stationary action, not from an extremal
principle.  As exemplified by Maxwell theory, this implies
conservation laws that include energy-momentum transport as well as
local conservation.
\par Since fields independent of $\Phi$ determine finite 
energy-momentum tensor $\Theta_m^{\mu\nu}$, the gravitational equation 
requires $X_\Phi^{\mu\nu}$ to be finite, regardless of any parameter 
values. This precludes spontaneous destabilization.  Given slowly 
varying $w^2$ and function $a(t)$ determined by the CHM Friedmann 
equation, Eq.(\ref{bphieq}) has a stable exact solution 
$b(t)\phi_0(t)$, which varies smoothly with parameters.
\par  The postulate of universal conformal symmetry for
primitive massless fields\cite{NES13} implies vanishing relativistic
trace for the energy-momentum tensor of each distinct field.
Vanishing total energy-momentum trace is retained for interacting
fields, subject to conservation laws.  This precludes constructing a
valid model of run-away expansion of the conformal Higgs scalar field
or the Friedmann scale function driven in the field equations 
by material fields of finite amplitude.
\par The conformal Higgs scalar field does not have a well-defined
mass, instead inducing dynamical $w^2$, which acts as a cosmological
constant determining dark energy.
Nearly constant ${\dot\phi}/\phi$ for redshifts $z\leq 1$ is shown here
to imply Higgs $\lambda<0$, consistent with a dynamical origin.  
Negative empirical $\lambda$ implies finite but `tachyonic'
mass for a scalar field fluctuation.  This does not support the
conventional concept of a massive Higgs particle\cite{NES10}.
A novel particle with Higgs properties may exist\cite{NESM4},
and could account for the sign and magnitude of $\lambda$.
Electroweak masses depend only on a stable Higgs scalar field of
finite amplitude, implied by the present analysis.

\section{Compatibility with conformal gravitation}
The postulate of universal local Weyl scaling (conformal) symmetry
requires modifying both general relativity and the Higgs scalar
field model.  No new fields are assumed.
Conformal gravity (CG) is confirmed by a fit to
rotation data for 138 galaxies\cite{MAN06,MAO11}.
The conformal Higgs model (CHM) acquires a gravitational
effect that fits observed Hubble expansion for redshifts
$z\leq 1$ (7.33 Gyr) accurately with only one free parameter.
Neither model requires dark matter.
Recent criticism of CG is resolved\cite{NESM5}.
Consistency is assured if both conformal models are implemented.
Nonclassical CG acceleration parameter $\gamma$ is determined by the
CHM.  A recently established empirical relationship between classical
and nonclassical galactic radial acceleration\cite{MLS16} requires 
nonclassical acceleration to be independent of galactic mass. Conformal
theory is found to be consistent with this and with the $v^4$ baryonic
Tully-Fisher relation for orbital rotation velocities\cite{NESM7,NESM3}.
Vanishing of centripetal acceleration outside a halo boundary,
a unique implication of conformal theory, is confirmed\cite{NESM5}.

\section{Conclusions}
\par These developments  modify standard cosmology qualitatively,
justifying persistent positive acceleration.  Acceleration history is
model-dependent. This may account for the current discrepancy in values
of the Hubble constant\cite{PLC15,RMH16}. Integration of the standard
Friedmann equation back to the CMB would be inaccurate if conformal
theory is correct.
\par The present analysis indicates that conformal theory can explain 
the existence and magnitude of dark energy.  This removes the huge 
disparity of parameters relevant to cosmological and elementary particle
phenomena.  Small parameter $\hbar^2w^2=1.53\times 10^{-66}eV^2$ is 
determined by a cosmological time derivative.  
$\hbar^2\phi^2(t_0)=3.03\times 10^{22}eV^2$ is a ratio of 
two small parameters, approximately $\half \hbar^2w^2/|\lambda|$.   
The underlying scale parameter is the Hubble constant. 
\par The present demonstration of consistency of the conformal Higgs
model with observed Hubble expansion has many practical implications.
It validates an unconventional cosmic history\cite{NESM1}. 
In particular, the non-Newtonian gravitational coupling parameter due 
to the Higgs scalar field has opposite sign to standard theory.  This 
implies accelerated initial expansion due to mass and radiation density,
a form of spontaneous `big bang' not requiring any ad hoc new theory.  
The CHM Friedmann equation,
omitting $\Omega_k$ and $\Omega_m$, is valid for $t\to\infty$,
implying cosmic expansion until a turnaround to contraction after an  
interval comparable to Hubble time unit $1/H_0$.
The CHM Friedmann acceleration weight is positive from the onset
of expansion, not at all a recent phenomenon, changing sign only as 
$t-t_0$ approaches Hubble time. 
\par An accurate solution for the early universe should include time
variation of electroweak parameters that depend on $\phi$,
including the Fermi constant for $\beta$ decay. 
\par The author is grateful to colleagues John Baglin and Barbara Jones 
for helpful comments.

\end{document}